\documentclass[12pt,letterpaper]{article}
\pdfoutput=1
\usepackage[utf8]{inputenc}
\usepackage{amsmath}
\usepackage{slashed}
\usepackage{amssymb}
\usepackage{amsthm}
\usepackage{tikz}
\usepackage{xcolor}
\usepackage{enumitem}
\usepackage[colorinlistoftodos]{todonotes}
\definecolor{nicered}{rgb}{0.7,0.1,0.1}
\definecolor{nicegreen}{rgb}{0.1,0.5,0.1}
\usepackage[colorlinks=true,citecolor= nicegreen,linkcolor=nicered]{hyperref}
\usepackage{graphicx}
\graphicspath{{figures/}}
\usepackage{siunitx} 
\usepackage[numbers,sort&compress]{natbib}
\usepackage{cancel}
\usepackage[colorinlistoftodos]{todonotes}

\usepackage{color}
\usepackage{soul}

\title{Vector Boson Fusion in the Inert Doublet Model}

\author{Bhaskar Dutta\footnote{\href{mailto:dutta@physics.tamu.edu}{dutta@physics.tamu.edu}}\\
\textit{\small  Mitchell Institute for Fundamental Physics and Astronomy,}\\ \textit{\small Department of Physics and Astronomy,
Texas A\&M University, College Station, TX 77843, USA}\\
[4mm]
Guillermo Palacio\footnote{\href{galberto.palacio@udea.edu.co}{galberto.palacio@udea.edu.co}}, Diego Restrepo\footnote{\href{mailto:restrepo@udea.edu.co}{restrepo@udea.edu.co}}\\
\textit{\small  Instituto de Física, Universidad de Antioquia, Calle 70 No. 52-21, Medellín, Colombia}\\
[4mm]
José D. Ruiz-Álvarez\footnote{\href{
jose.ruiz@cern.ch }{
jose.ruiz@cern.ch}}\\
\textit{\small Departamento de Física, Universidad de Los Andes, Código Postal 111711, Bogotá, Colombia}
}
\date{\small Month NN, YYYY}
\begin{document}
\maketitle
\begin{tikzpicture}[overlay, remember picture]
\path (current page.north east) ++(-1,-1) node[below left] {MI-TH-1767};
\end{tikzpicture}

\begin{abstract}
In this paper we probe inert Higgs doublet model at the LHC using Vector Boson Fusion (VBF) search strategy. We optimize the selection cuts and investigate the parameter space of the  model and we show that  the  VBF search has a better reach when compared with the monojet searches. We also investigate the Drell-Yan type cuts and show that they can be important  for smaller charged Higgs masses. We determine the $3\sigma$ reach for the parameter space using these optimized cuts for a luminosity of 3000 fb$^{-1}$. 
\end{abstract}
\section{Introduction}
The particle physics origin of the 27\% of the universe is still unknown.  The results from the direct, indirect detections and the collider experiment are investigating  particle physics models which provide a dark matter candidate. The null results so far from these experiments have already ruled out many models. However, many models still exist with large regions of allowed parameter space.  Since the LHC is ongoing, it will be crucial to come up with strategies to investigate  the parameter spaces of these models to the maximum extent.  In this paper, our main focus is to use the LHC searches to investigate  one simple dark matter model by developing a search strategy and compare with the existing search strategies.

  One of the simplest models which provides a dark matter (DM) candidate  is the Inert Doublet Model (IDM)~\cite{Deshpande:1977rw,Barbieri:2006dq},
where an additional scalar doublet of $SU(2)_L$ odd under a global $Z_2$ is added to the Standard Model (SM). 
The lightest  neutral component of the  scalar doublet acts as a dark matter candidate. 
Recent accounts of the model are given in~\cite{Eiteneuer:2017hoh,Belyaev:2016lok,vonderPahlen:2016cbw,Garcia-Cely:2015khw,Queiroz:2015utg,Blinov:2015qva}, and further references are avaliable therein.
In particular, regarding the predictions of the model at colliders, the Drell-Yann (D-Y) production with
lepton plus transverse missing energy signals at LHC have been extensively studied in the IDM as 
in~\cite{Hashemi:2016wup,Datta:2016nfz,Blinov:2015qva,Belanger:2015kga,Gustafsson:2012aj,Miao:2010rg,Dolle:2009ft,Cao:2007rm}.
We plan to investigate this model in this paper utilizing the Vector Boson Fusion (VBF) search strategy. Since the dark matter candidate of this model has weak charges,  $W$, $Z$ fusions are useful to produce these particles at the LHC. We will optimize the VBF cuts to improve the signal to the background ratio where the SM background mostly arises from $Z$+ jets. Utilizing the optimized cuts,  we will show the reach of the LHC for the parameter space of the model in the ongoing and future runs. We will also compare this analysis with the existing search strategies.

We will investigae  the LHC reach of the parameter space of the model without any restrictions arising from annihilation rate, direct and indirect detections since the correlations among these results to constrain the parameter space for the LHC requires many assumptions. For example, the annihilation rate constraint arising from the DM content requires a knowledge of the history of the universe prior to big bang nucleosynthesis (BBN) which is unconstrained. The origin of the dark matter content, e.g., may be due to thermal, non-thermal~\cite{Barrow:1982ei, Kamionkowski:1990ni,Moroi:1999zb,Fujii:2001xp, Gelmini:2006pw,Kitano:2008tk,Dutta:2009uf}, non-standard cosmology (where the expansion rate can be different compared to the standard cosmology prior to the BBN)~\cite{Catena:2004ba,Gelmini:2008sh,Lahanas:2006hf,Meehan:2015cna,Dutta:2016htz}. Consequently, the  annihilation rate corresponding to the 27\% can be very different compared to the thermal dark matter in standard cosmology and in some non-thermal scenarios, the dark matter content may not be related to any annihilation rate at all.  We, therefore, plan to search for this model at the LHC without showing any preference for a particular cosmological scenario. Further, there can be multiple DM candidates (e.g., axion and DM from the  inert doublet model) and in such scenarios~\cite{Baer:2015tva,Aparicio:2015sda},  the direct and indirect detection cross-sections should be reduced by $R$ and $R^2$ respectively with $R\equiv {{\Omega h^2}/{0.12}}$. From all these considerations, it appears that the search at the LHC  should be strategized without applying restrictions arising from the  thermal annihilation rate and the direct and indirect detection constraints. If  the signals  from  a particular physics model which possess a DM candidate are discovered at the LHC, we would not only be able to establish that model but it also would give us an opportunity to investigate  the cosmology in the Pre-BBN era. 

Following this strategy,  the Monojet final state has been used effectively to search for parameter space of this model in ref.~\cite{Belyaev:2016lok}. In this paper, we  first optimize VBF cuts to search for the DM candidate in IDM. We will show that the VBF reach is better than the monojet final state. We will also show the parameter space where the D-Y type cuts can be important.

The paper is organized as follows. In section~\ref{sec:model}, we discuss the model. In section~\ref{sec:search}, we discuss the VBF signatures and develop VBF cuts. In section~\ref{sec:results}, we discuss the parameter space reach for this model at the LHC and we conclude in section~\ref{sec:conclusions}.

\section{The model}
\label{sec:model}

The Inert Doublet Model (IDM) is a minimal extension of the SM,
where an additional $SU(2)_L$ scalar doublet, $\Phi$, odd under a $Z_2$
discrete symmetry is added. The lightest  neutral component of the scalar
doublet is the dark matter candidate.

The Lagrangian of the model is given by:
\begin{align}
\label{eq:0001}
\mathcal{L} =\mathcal{L}_{\text{SM}}+ (D^{\mu}\Phi)^{\dagger} D_{\mu}\Phi -V(\Phi),
\end{align}
where
\begin{align}
\label{eq:0002}
V(\Phi) =& \mu_2^{2} \Phi^{\dagger}\Phi + \lambda_2 (\Phi^{\dagger}\Phi)^{2} + 
\lambda_3 (H^{\dagger}H)(\Phi^{\dagger}\Phi)  + \lambda_4 (H^{\dagger}\Phi)(\Phi^{\dagger}H) + \dfrac{\lambda_5}{2} \Big [ (\Phi^{\dagger}H)^2 + \text{h.c.} \Big ].
\end{align} 
$H$ stands for the Higgs doublet in the SM, which 
acquires a vacuum expectation value (VEV) $v=246.2$~GeV. We define
\begin{align}
\lambda_L=\frac{\lambda_3+\lambda_4+\lambda_5}{2}\,.
\end{align}
which controls the interactions though the Higgs portal. The components of the scalar doublets are defined as

\begin{align}
\label{eq:0003}
\Phi=\begin{pmatrix}
 H^{+} \\
\dfrac{1}{\sqrt{2}}(H^{0} + i A^{0}) 
\end{pmatrix}, \hspace{2cm}
H=\begin{pmatrix}
 G^{+} \\
\dfrac{1}{\sqrt{2}}(v + h^{0} + i G^{0}) 
\end{pmatrix}.
\end{align}
The fields are written in the canonical form. $H^{0}$, $A^{0}$ and $H^{+}$  are the CP even, CP odd and
charged scalar of the additional scalar doublet.  $G^{+}$ and
$G^{0}$ are 
 the Goldstone bosons of the the $SU(2)_L \times U(1)_Y$ electroweak symmetry,

\section{Search for VBF signature at the LHC }
\label{sec:search}
The IDM can be explored in the current and futures runs of the 
large hadron collider (LHC) with a
center of mass energy $\sqrt{s}=13\ \text{TeV}$. Since the DM candidate, $H_0$ has weak charges, it can be produced via a VBF strategy~\cite{Delannoy:2013ata} at the LHC. The VBF search for the DM  investigate missing energy in the association of two high $E_T$ jet with large $|\Delta\eta|$ between the jets in the final state.

\begin{figure}[h]
\begin{centering}
\def\sep{\hspace{0.3cm}}
\def\scl{0.23}
\includegraphics[scale=\scl]{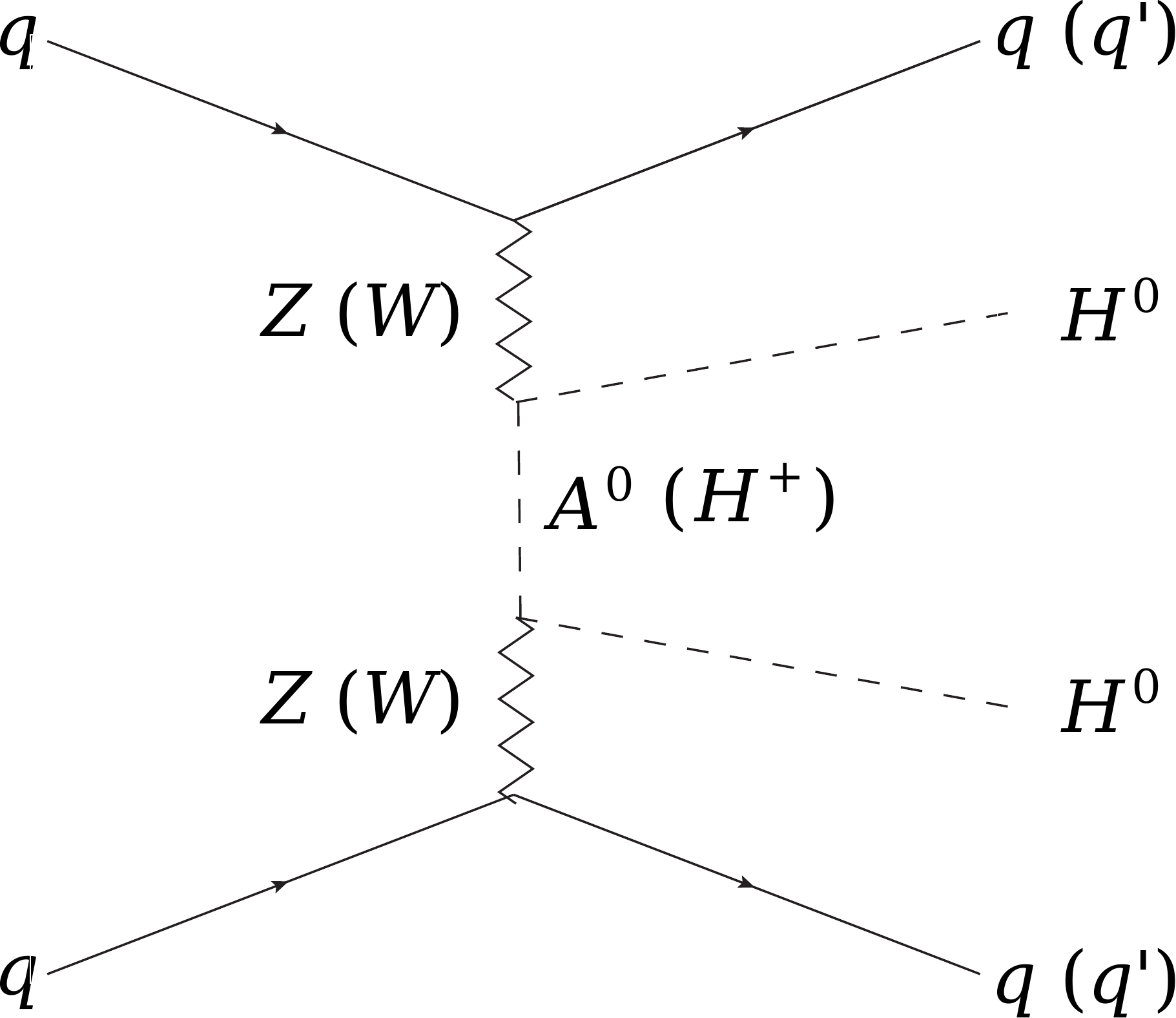}\sep
\includegraphics[scale=\scl]{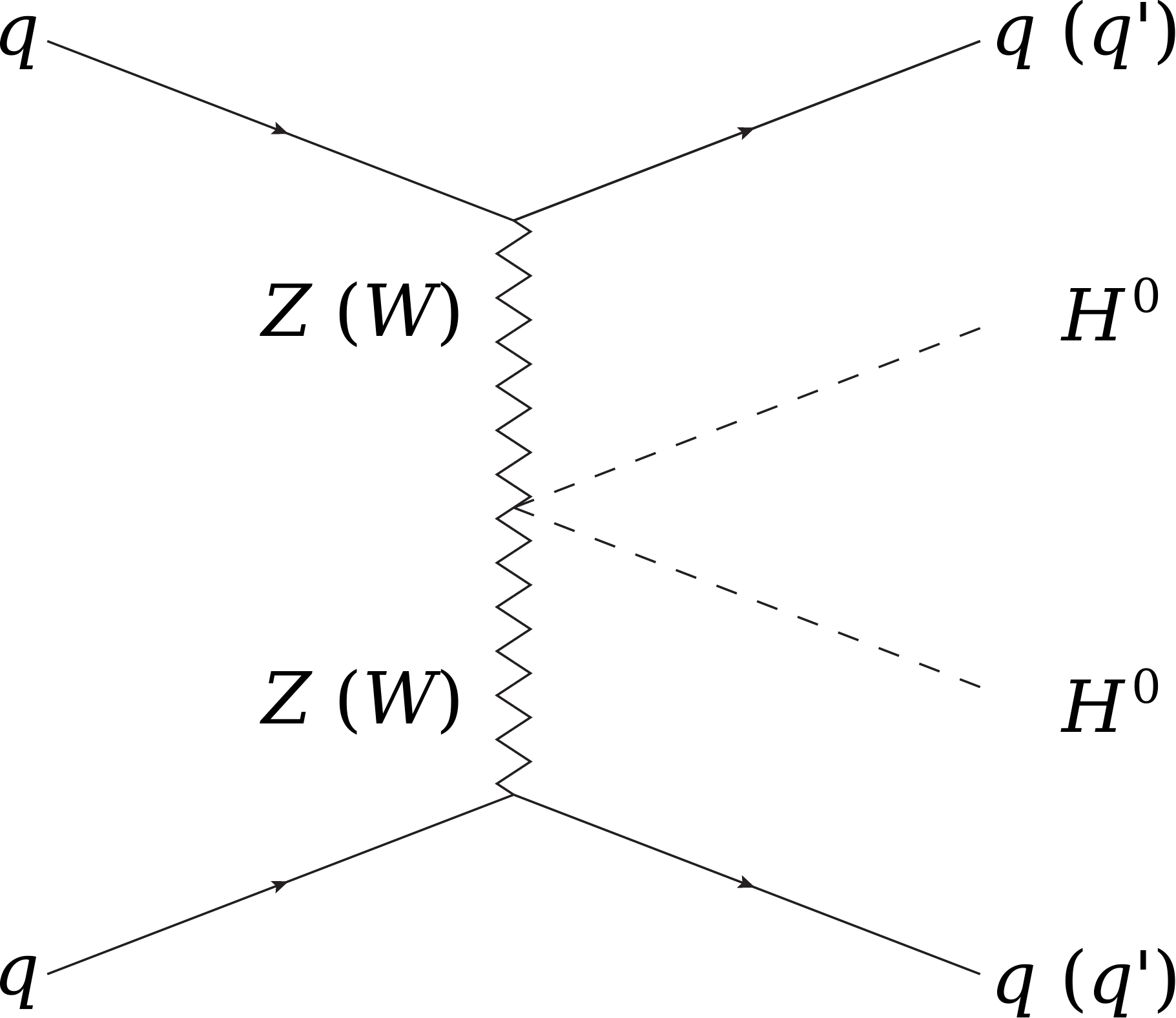}\sep
\includegraphics[scale=\scl]{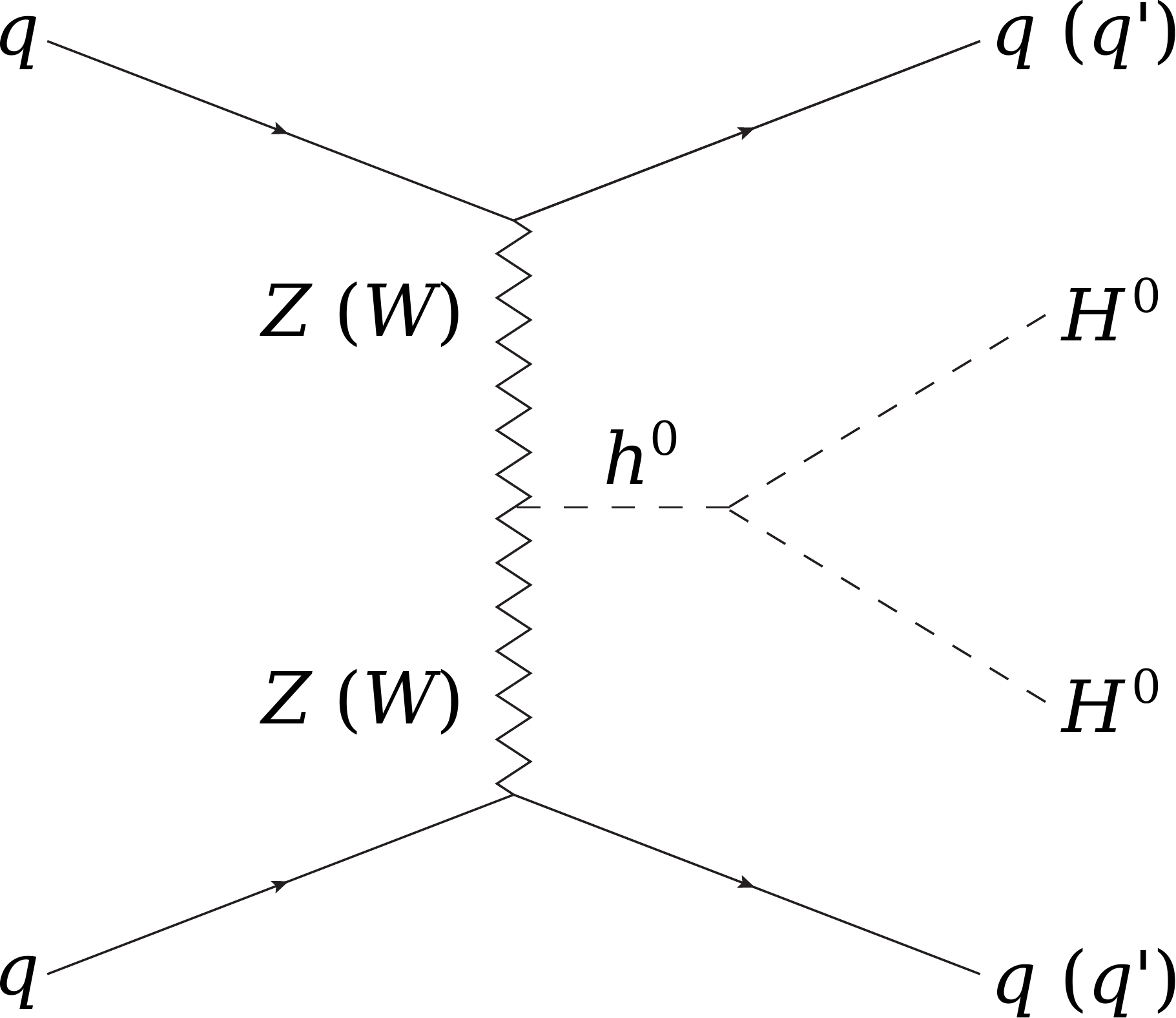}\sep
\includegraphics[scale=\scl]{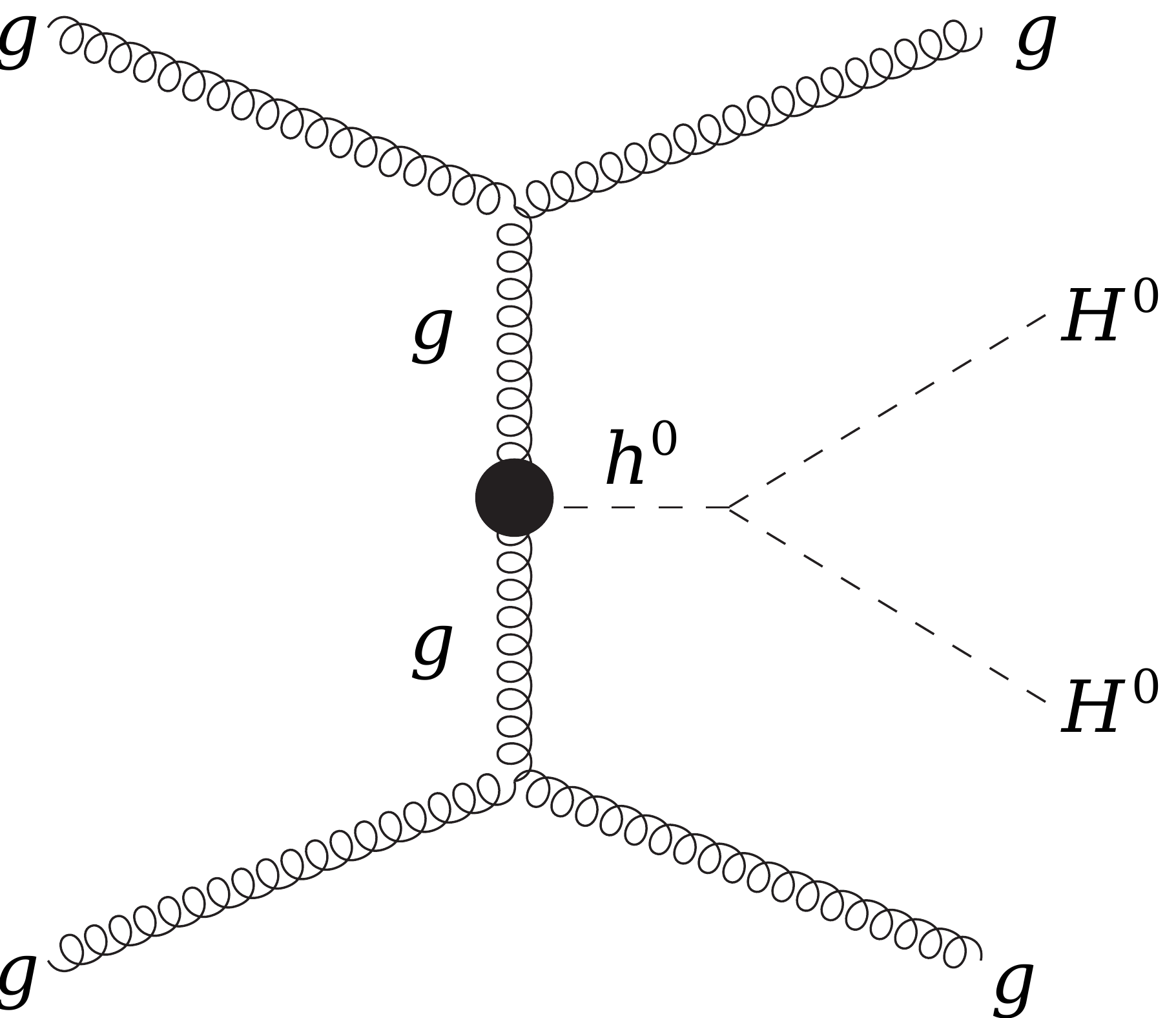}
\caption{\label{fig:FD} 
Feynman diagrams which contribute to $ p p \to H^{0} H^{0} j j $ in the IDM
}
\end{centering}
\end{figure}

The important contributions to the  VBF  total cross-section for this model
are displayed in figure~\ref{fig:FD}. The last two diagrams are only important for large values of $\lambda_L$, while the first two diagrams can have large destructive interferences for small values of $\lambda_L$ and not too large splitting between the set of inert scalar masses.  The blob in the gluon diagram denotes the effective coupling between the gluons and the SM Higgs.

\begin{figure}
  \centering
  \includegraphics[scale=0.5]{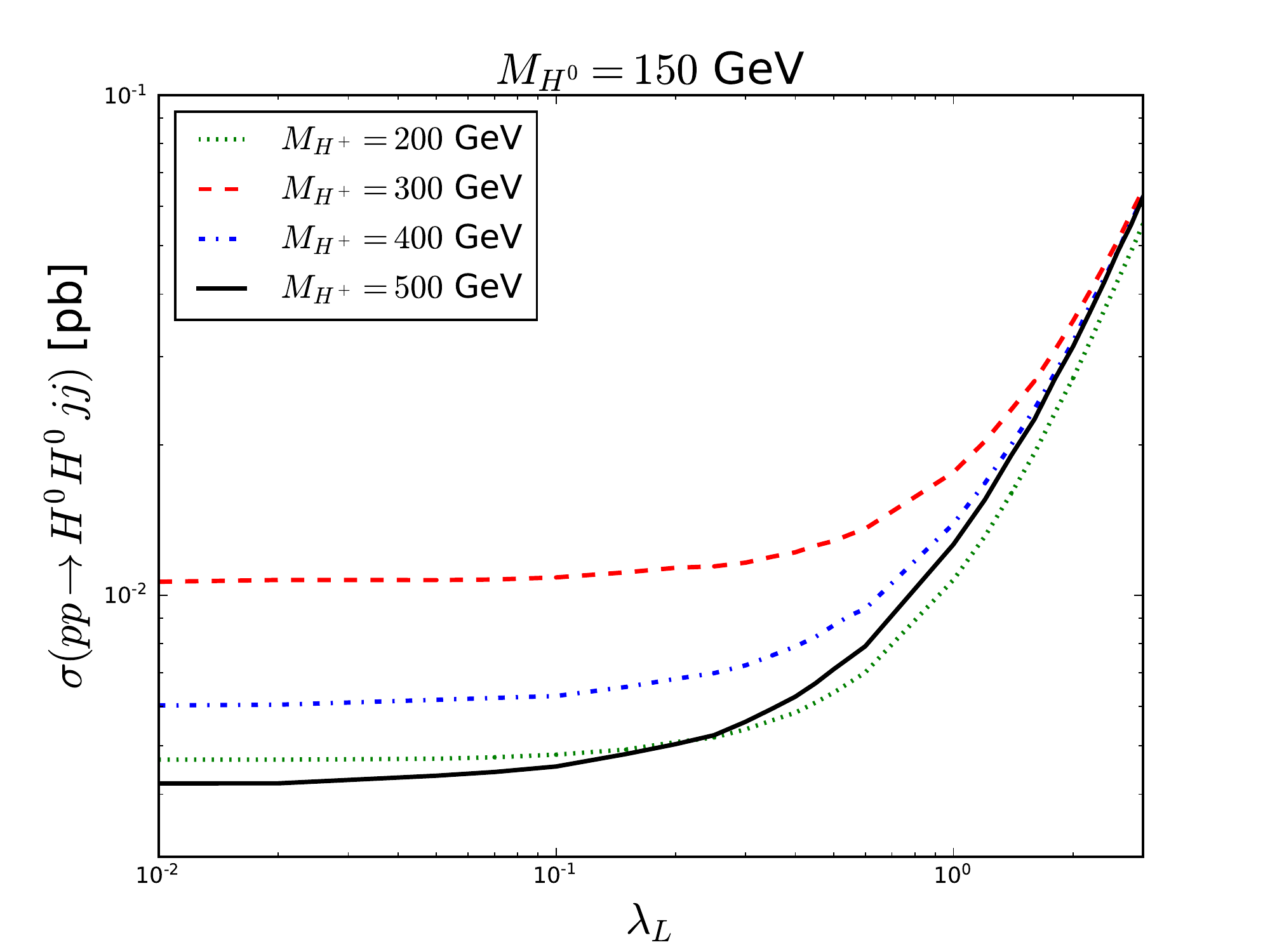}
  \caption{Cross section for the process $pp\to H^0 H^0 \,j j$  as a function of $\lambda_L$ for $M_{H^0}=150\ \text{GeV}$, and several values of $M_{A^0}=M_{H^+}$.}
  \label{fig:csls}
\end{figure}

The inclusive cross section for the process $pp\to H^0 H^0 \,j j$ for a fixed dark matter mass of $150$~GeV is displayed in figure~\ref{fig:csls} as a function of $\lambda_L$ for several values of $M_{A^0}=M_{H^+}$. Note that for specific values of $M_{A^0}=M_{H^+}$, the Drell-Yann production of inert scalars for  small $\lambda_L$ can be enhanced, because of resonant production of gauge bosons which give rise to the two jets.

The VBF topology relies in a set of characteristic of the events from the point of view of a detector as the ATLAS or CMS experiments. The two jets produced from such signature are located in different hemispheres of the detector, which means that $\eta(j_{1})\times\eta(j_{2})<0$. Additionally, these two jets are also well separated in the pseudorapidity. We expect then that the two jets from backgrounds faking the VBF topology have smaller separation in $\eta$ than the two jets from our signal. Finally, a key characteristic of the two jets from the VBF topology is that their invariant mass is larger than for a couple of non-VBF jets.
Some contributions to the VBF signal have been considered in \cite{Poulose:2016lvz} which does not include some of the essential VBF cuts, e.g., large $|\Delta\eta|$, large $M_{j_1 j_2}$ etc. Our cuts originate from   the experimental searches as mentioned above~\cite{Khachatryan:2016mbu, Aad:2015txa} which we optimize further. Consequently, we obtain much larger reach for the parameter space. 

All these VBF characteristics significantly reduce backgrounds and gives us a different set of background than monojet searches. Our main  backgrounds are $Z\to\nu\bar{\nu}$+jets and $W^{+/-}\to l\bar{\nu}$+jets, where the lepton is missed by the detector reconstruction (for example if it is produced outside the experiment acceptance or fails isolation criteria). $Z\to\nu\bar{\nu}$+jets background will be referred in the following as simply $Z$+jets. The QCD contribution to our background expectations is very small, and we consider it negligible for simplicity of this work.

To design our analysis we have used MC simulations of $Z$+jets and the signal. We have used MadGraph~\cite{Alwall:2014hca} to simulate the partonic process while Pythia 8~\cite{Sjostrand:2014zea} for the hadronization and showering. Finally, we have processed our samples with Delphes~\cite{deFavereau:2013fsa} to simulate a detector response. We have used default configurations from the packages and we have worked specifically with the CMS experiment simulation done by Delphes. We do not expect significant differences to our conclusions by switching the detector simulation to an ATLAS-like configuration. We have used AK4 (Anti-kt algorithm with $R=0.4$) jets that are reconstructed with fastjet package~\cite{Cacciari:2011ma}. Jets were reconstructed in a rapidity acceptance of $|\eta(j)|<5$. For the simulation of the signal we have used the IDM model implementation~\cite{Belanger:2015kga,Goudelis:2013uca} available in the FeynRules~\cite{Alloul:2013bka} models database.

Based on the analysis presented in~\cite{Khachatryan:2016mbu}, we assume that the $W^{+/-}$+jets background is kinematically similar to the $Z$+jets background. and that after full selection our background contribution will be composed 70\% by $Z$-jets events and 30\% by $W^{+-}$+jets. Moreover, as we just consider the significance
\begin{align}
\sigma\equiv \frac{S}{\sqrt{S+B}} \,,
\end{align}
as figure of merit over the total number of events to determine the goodness of the selection, we are not affected by potential small kinematic differences between our two main backgrounds. 

For the analysis we propose we relied in the following set of variables:
\begin{itemize}
\item $p_{T}^{\text{miss}}=-|\sum_{i=0}^{N(j)} \vec{p}_{T}(j_{i})|$ denoted in the literature as transverse missing energy.
\item $N(j)$ the jet multiplicity.
\item The two leading jets $p_{T}$, $p_{T}(j_{1})$ and $p_{T}(j_{2})$.
\item $\eta(j_{1})\times\eta(j_{2})$
\item $|\Delta\eta(j_{1},j_{2})|$
\item $M(j_{1},j_{2})$, the invariant mass of the two leading jets.
\end{itemize}

The selection was followed having the greatest significance in the
order of the variables they have been cited. $N(j)$ was fixed to be
greater than 1 and $\eta(j_{1})\times\eta(j_{2})<0$. The signal
samples used for the optimization was produced with 
$M_{H^{0}}=65$~GeV,  $M_{H^{+}}=M_{A^{0}}=750$~GeV and $\lambda_{L}=0.2$,
but it has been checked that changing the $\lambda_{L}$ parameter we do not gain
re-optimizing the selection. 


The set of cuts that drives the analysis to the greatest significance is:

\begin{enumerate}[label=Cut \arabic*,labelindent=\parindent,leftmargin=*]
\item:  $p_{T}^{\text{miss}}>180$~GeV\label{cut:1}
\item:  $N(j)\ge 2$ \label{cut:2}
\item:  $p_{T}(j_{1})>100$~GeV \label{cut:3}
\item:  $p_{T}(j_{2})>50$~GeV \label{cut:4}
\item:  $\eta(j_{1})\times\eta(j_{2})<0$ \label{cut:5}
\item:  $|\Delta\eta(j_{1},j_{2})|>4.2$ \label{cut:6}
\item:  $M(j_{1},j_{2})>1$~TeV\label{cut:7}
\end{enumerate}

\begin{figure}
  \centering
  \includegraphics[scale=0.5]{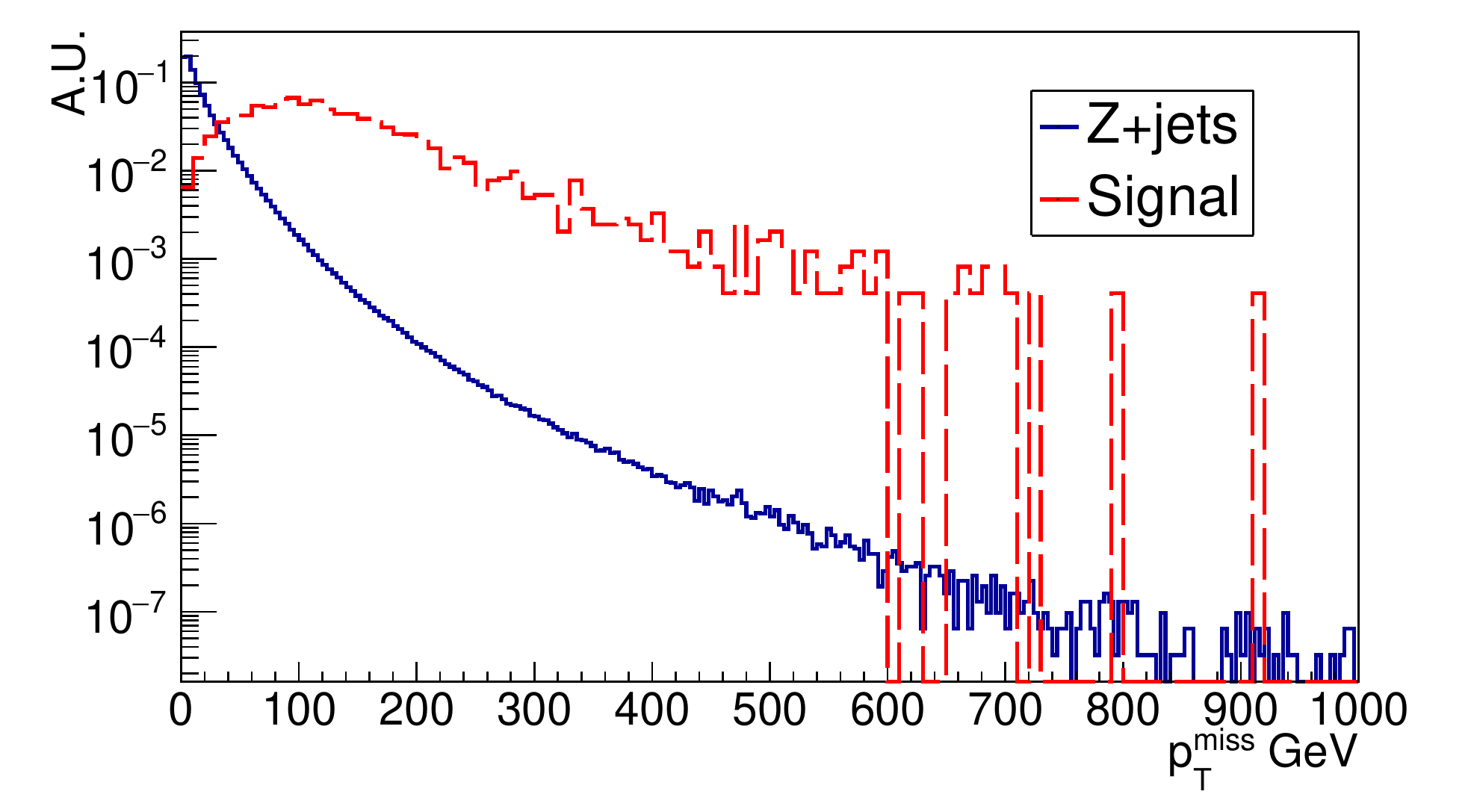} 
  \caption{$p_{T}^{miss}$ signal and $Z$+jets background distributions before analysis selection. Both distributions have been normalized to unity.}
  \label{fig:MET}
\end{figure}

The selection on the missing energy was chosen quite high because this variable is normally used for the trigger in the experiments for dark matter searches. It would be a good improvement for this search if this threshold could be lowered down as much as possible in the triggers used by the experiments. A comparison between signal and the main background on this variable before any cut can be found in figure~\ref{fig:MET}. It can be seen that our signal is expected to have greater missing transverse momentum than the main background. 

\begin{figure}
  \centering
  \includegraphics[scale=0.5]{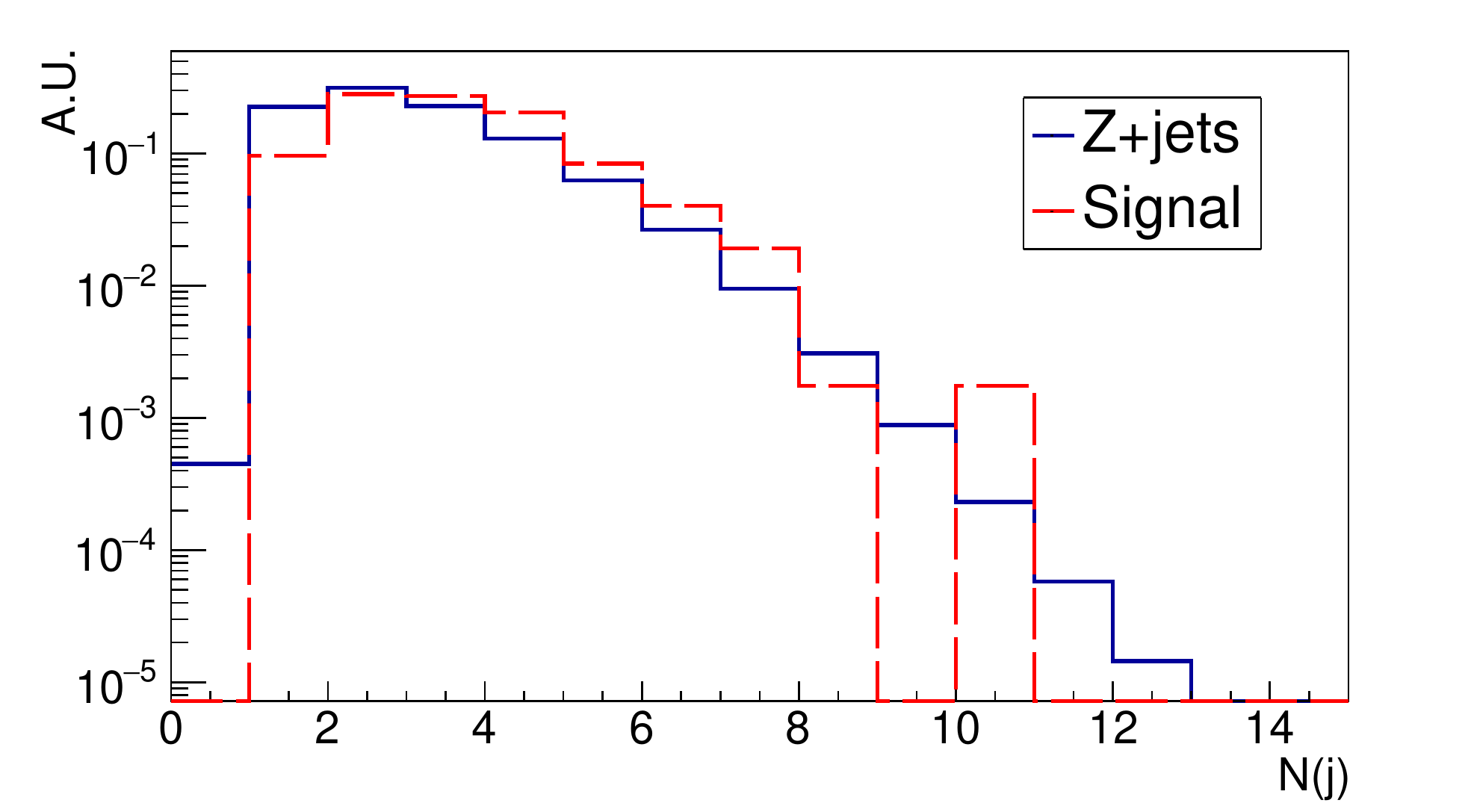} 
  \caption{$N(j)$ signal and $Z$+jets background distributions after cut 1. Both distributions have been normalized to unity.}
  \label{fig:NJ}
\end{figure}

\begin{figure}
  \centering
  \includegraphics[scale=0.4]{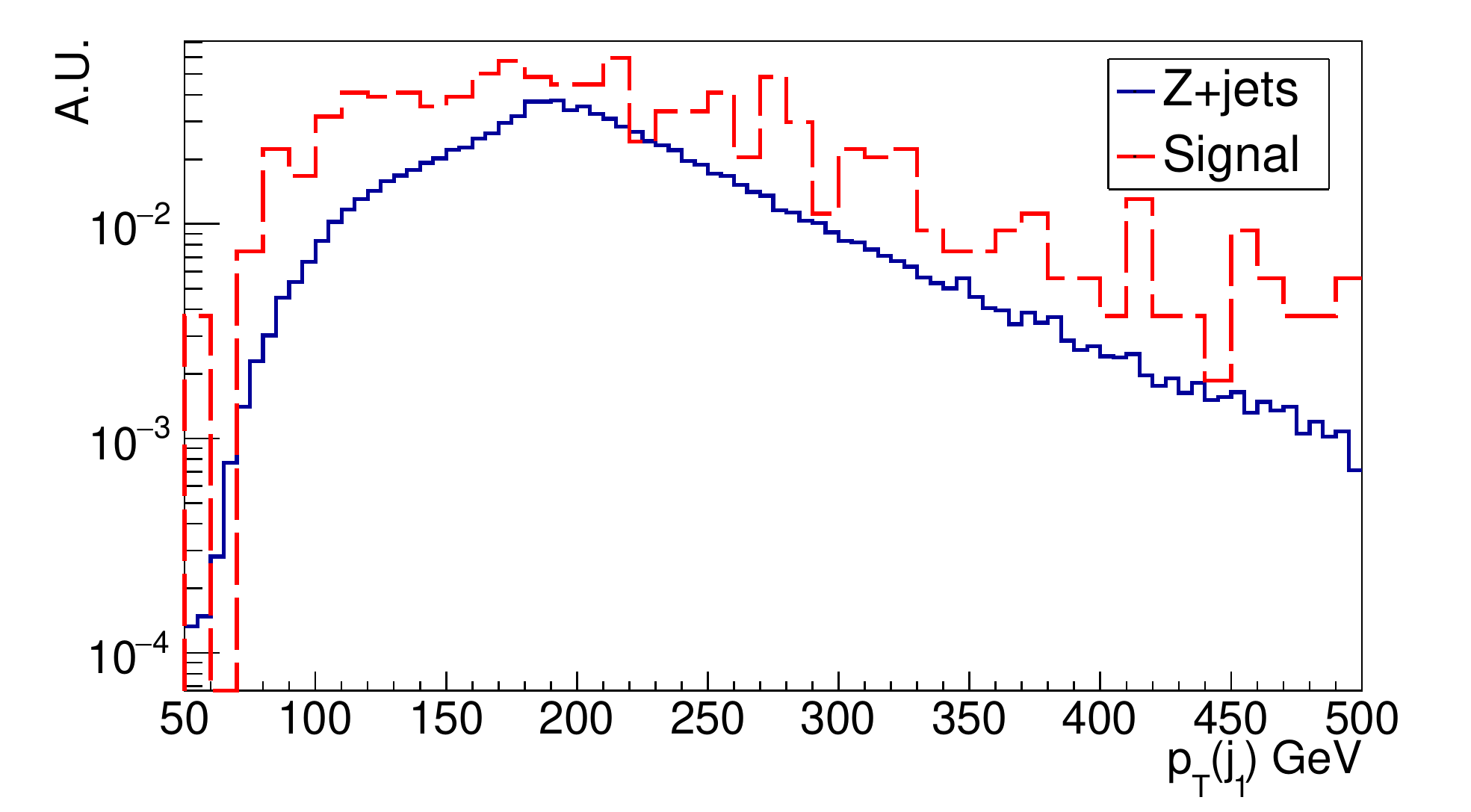} 
  \includegraphics[scale=0.4]{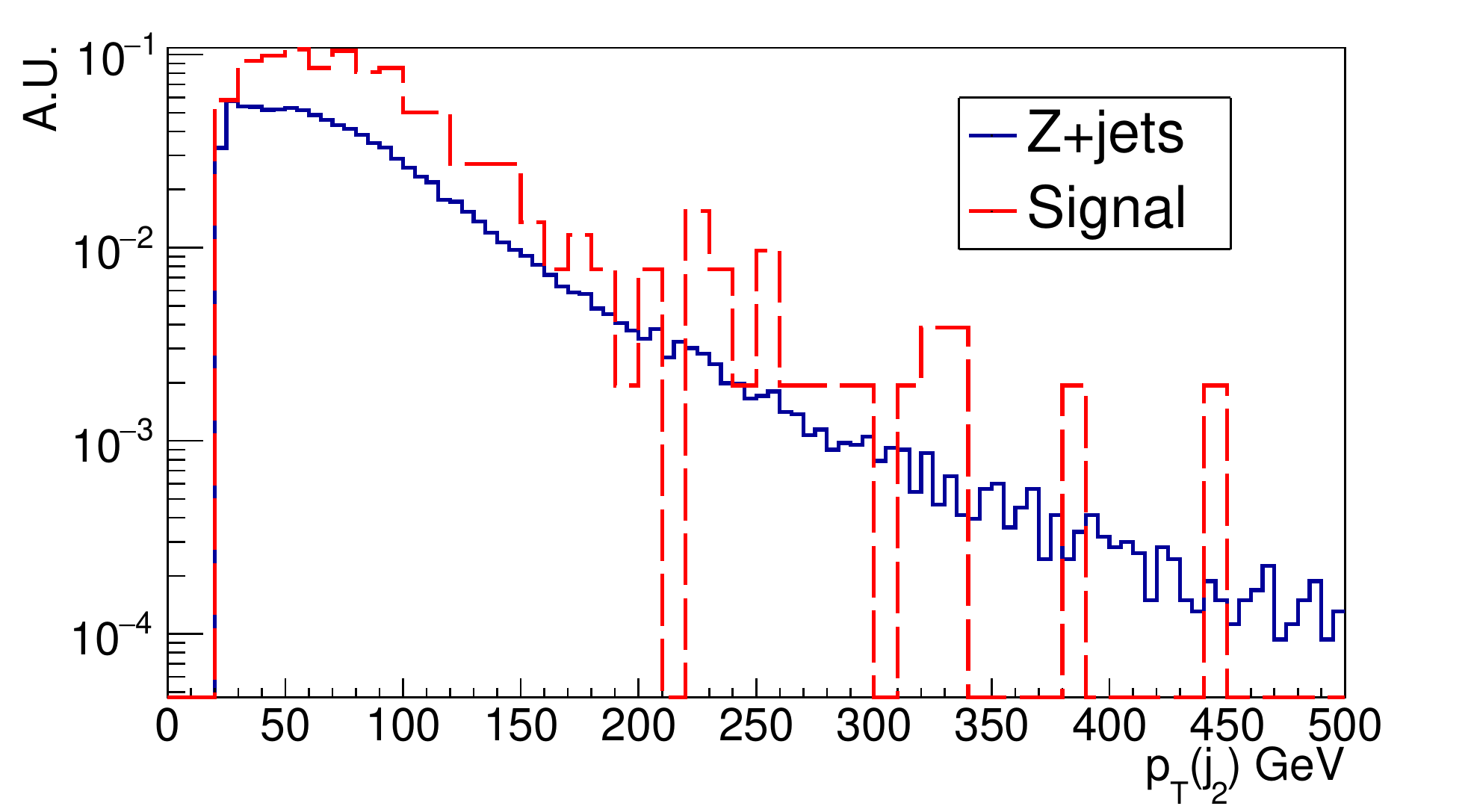} 
  \caption{$p_{T}(j_{1})$ (left) and $p_{T}(j_{2})$ (right) signal and $Z$+jets background distributions before analysis selection. Both distributions have been normalized to unity.}
  \label{fig:JPT}
\end{figure}

In figure~\ref{fig:NJ}  the jet multiplicity of our signal and main background after~\ref{cut:1} is shown. Figure~\ref{fig:JPT} shows the leading jets transverse momentum at the same stage of the selection and for the same samples. From the jets $p_{T}$ distributions it can be seen that however the main background tends to have quite energetic first leading jet and a gain in sensitivity can be achieved cutting the events at the lower tail of the distribution. For the sub-leading jet, the signal shows clearly more energetic jets than the main background.

\begin{figure}
  \centering
  \includegraphics[scale=0.5]{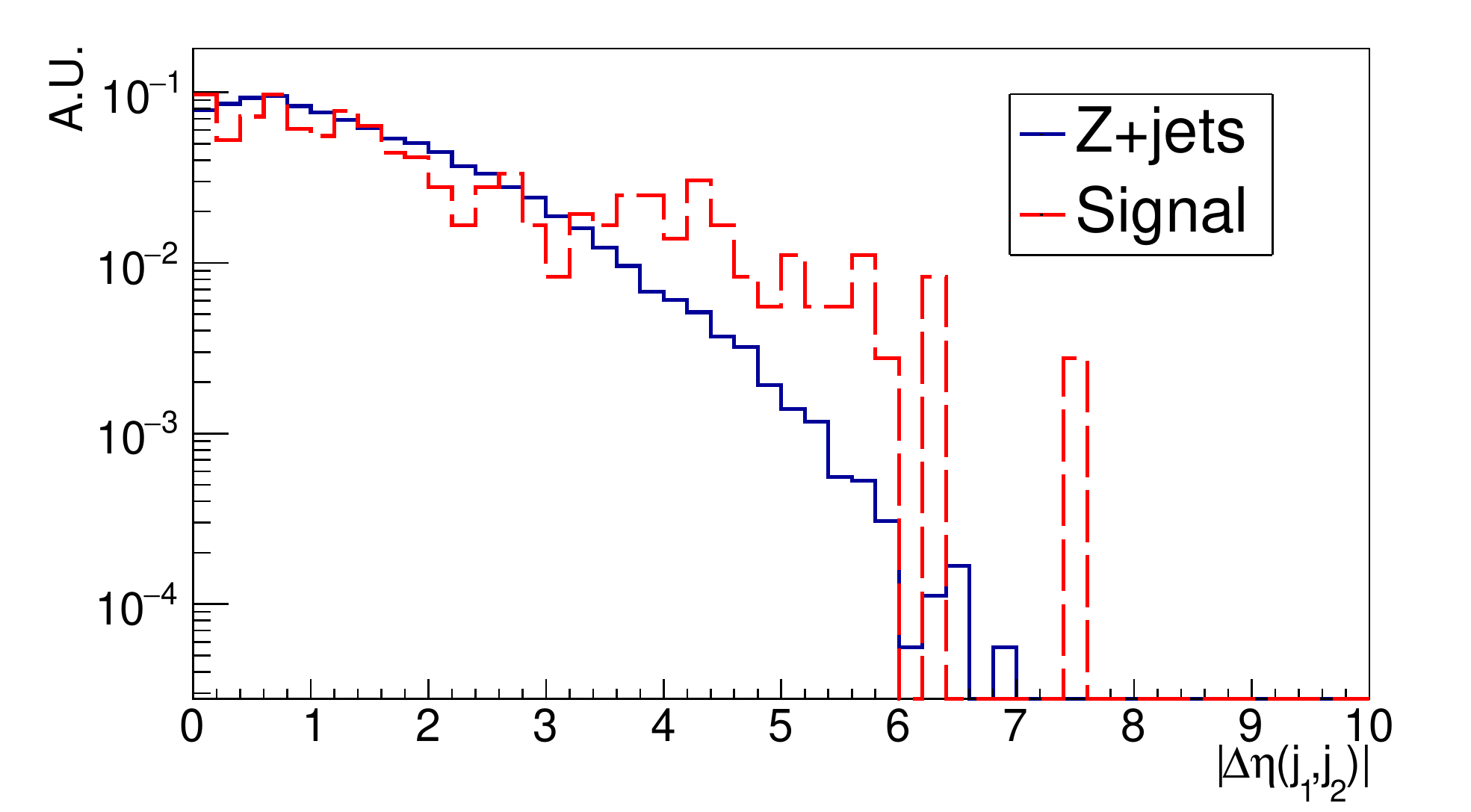} 
  \caption{$\Delta\eta(j_{1},j_{2})$ signal and $Z$+jets background distributions after cut 4. Both distributions have been normalized to unity.}
  \label{fig:Deta}
\end{figure}

The pseudorapidity separation of the two leading jets is shown in figure~\ref{fig:Deta} after~\ref{cut:4}. We can corroborate from all these cited distributions  that signal tends to have greatest separation between the two leading jets than the main background.

\begin{figure}
  \centering
  \includegraphics[scale=0.5]{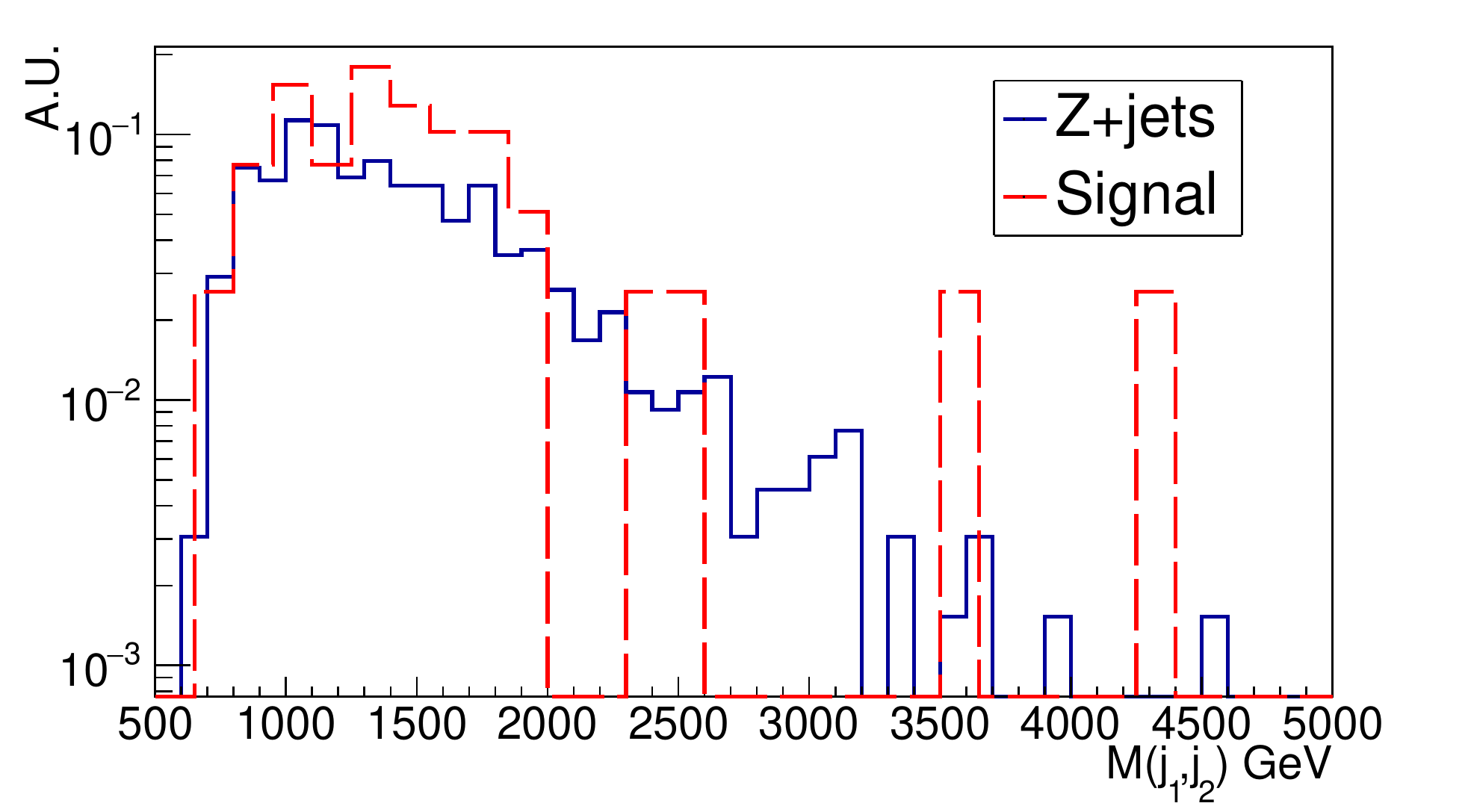} 
  \caption{$M(j_{1},j_{2})$ signal and $Z$+jets background distributions after cut 6. Both distributions have been normalized to unity.}
  \label{fig:Mjj}
\end{figure}

Finally, further differences between signal and the main background can be found in the invariant mass of the two leading jets. The distribution for signal and $Z$+jets background after cut 6 are shown in figure~\ref{fig:Mjj}. 

The efficiencies of each cut are displayed in table~\ref{tab:cutflow}.

\begin{table}[h]
\centering
\resizebox{\textwidth}{!}{
\begin{tabular}{|c|c|c|c|c|} 
 \hline
 & \multicolumn{2}{c|}{Efficiency per cut} & \multicolumn{2}{c|}{Cumulative efficiency} \\ \hline
Process & Signal & $Z$+jets & Signal & $Z$+jets \\ \hline \hline
Initial number of MC events & 2447 & 30996944 & 2447 & 30996944 \\ \hline
\ref{cut:1} & ($23.38 \pm 0.86$)\% & ($0.22 \pm 8\times10^{-4}$)\% & ($23.38 \pm 0.86$)\% & ($0.22 \pm 8\times10^{-4}$)\% \\ \hline
\ref{cut:4}& ($63.11 \pm 2.02$)\% & ($51.95 \pm 0.19$)\% & ($14.75 \pm 0.72$)\% & ($0.12 \pm 6\times10^{-4}$)\% \\ \hline
\ref{cut:6} & ($10.80 \pm 1.63$)\% & ($1.82 \pm 0.07$)\% & ($1.59 \pm 0.25$)\% & ($2.11\times10^{-3} \pm 8.3\times10^{-5}$)\% \\ \hline
\ref{cut:7} & ($84.62 \pm 5.78$)\% & ($82.57 \pm 1.48$)\% & ($1.35 \pm 0.23$)\% & ($1.74\times10^{-3} \pm 7.5\times10^{-5}$)\% \\ \hline
\end{tabular}
}
\caption{Efficiencies for signal and $Z$+jets background for different stages of the selection.}
\label{tab:cutflow}
\end{table}

In table~\ref{tab:cutflow} we have cited the efficiencies of the selection for a signal with $\lambda_{L}=0.2$. These efficiencies actually have a dependence on this parameter. When $\lambda_{L}$ is greater than 1, the first diagram displayed in figure~\ref{fig:FD} becomes subdominant, and therefore changing the selection efficiency. Therefore we have scanned $\lambda_{L}$ between 0.01 and 10, and we found efficiencies between 1 and 5\%. The efficiencies obtained from this scan have been used in the results section to calculate the significance as a function of $\lambda_{L}$. From table~\ref{tab:cutflow} we can see that \ref{cut:7} is not resulting in a strong increase in the selection discrimination power. However, this cut, inspired from experimental results~\cite{Khachatryan:2016mbu, Aad:2015txa}, would be more discriminant with a different technique to estimate backgrounds as with data-driven methods.

\section{Results}
\label{sec:results}
\begin{figure}
  \centering
  \includegraphics[scale=0.42]{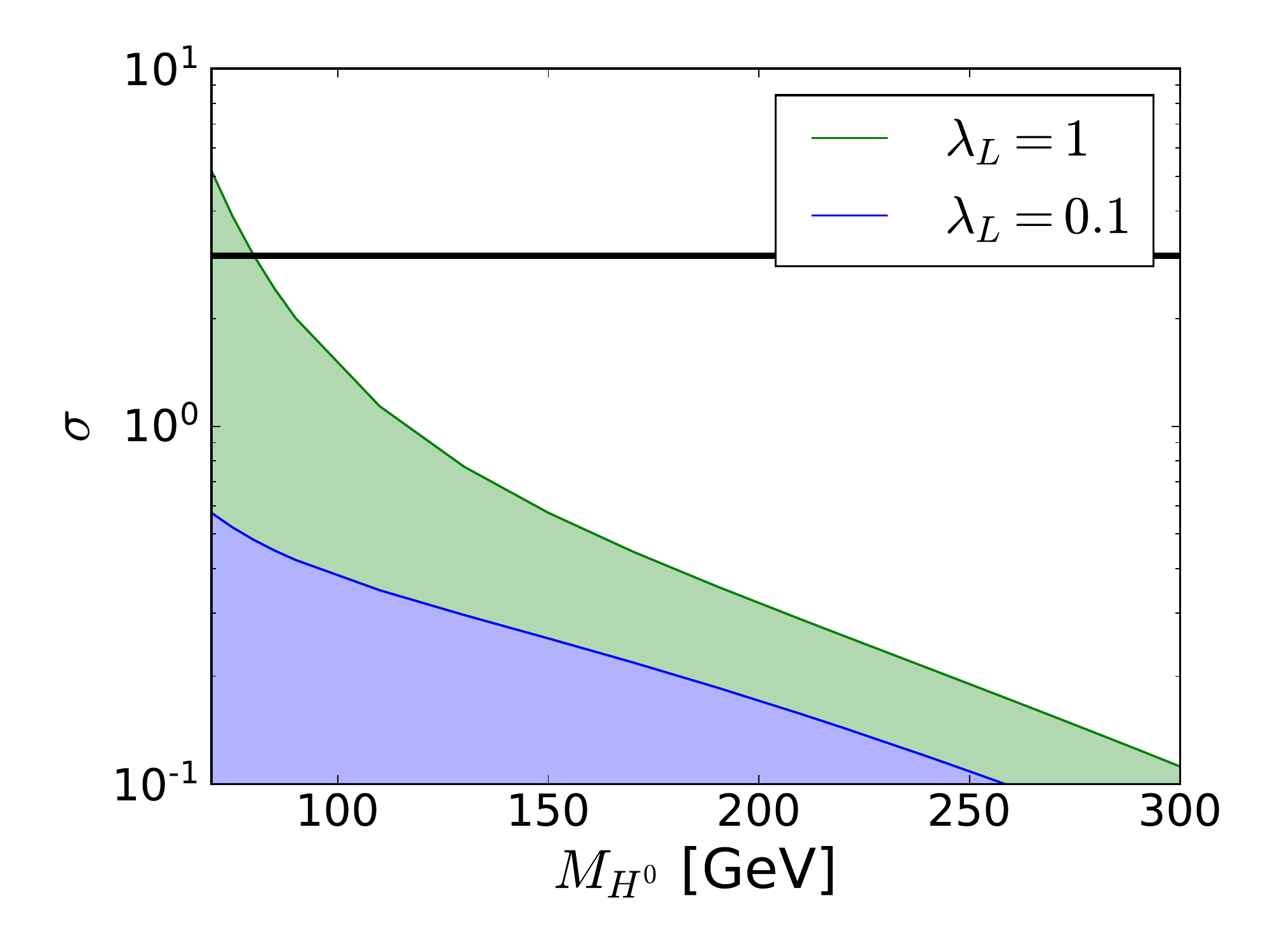} 
  \includegraphics[scale=0.43]{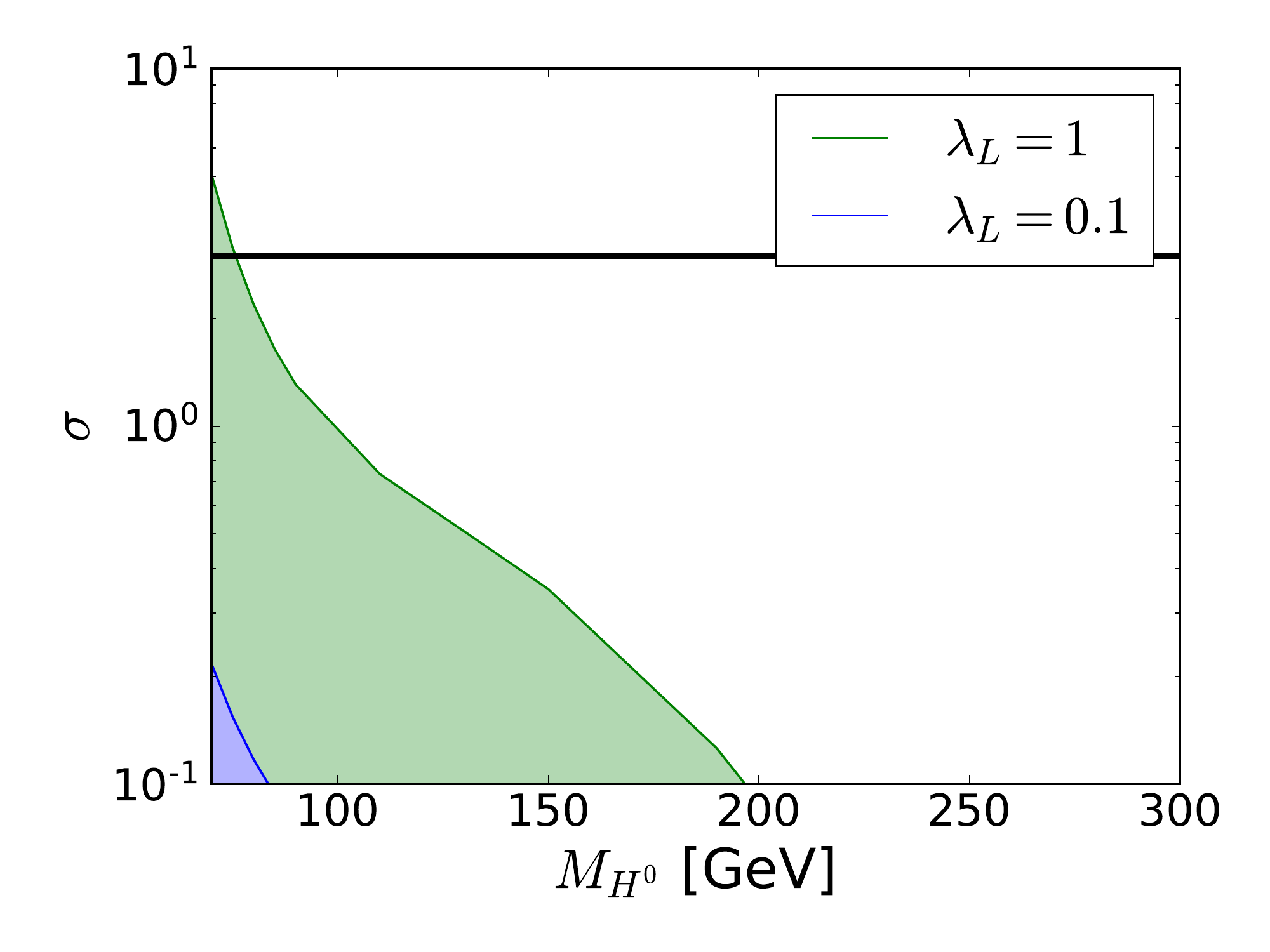}
  \caption{Significance for DM candidate discovery  for a luminosity of $30\ \text{fb}^{-1}$. The charged scalar mass is $750\ \text{GeV}$ ($250\ \text{GeV}$) in the left (right) panel }
  \label{fig:f1}
\end{figure}

\begin{figure}
  \centering
  \includegraphics[scale=0.43]{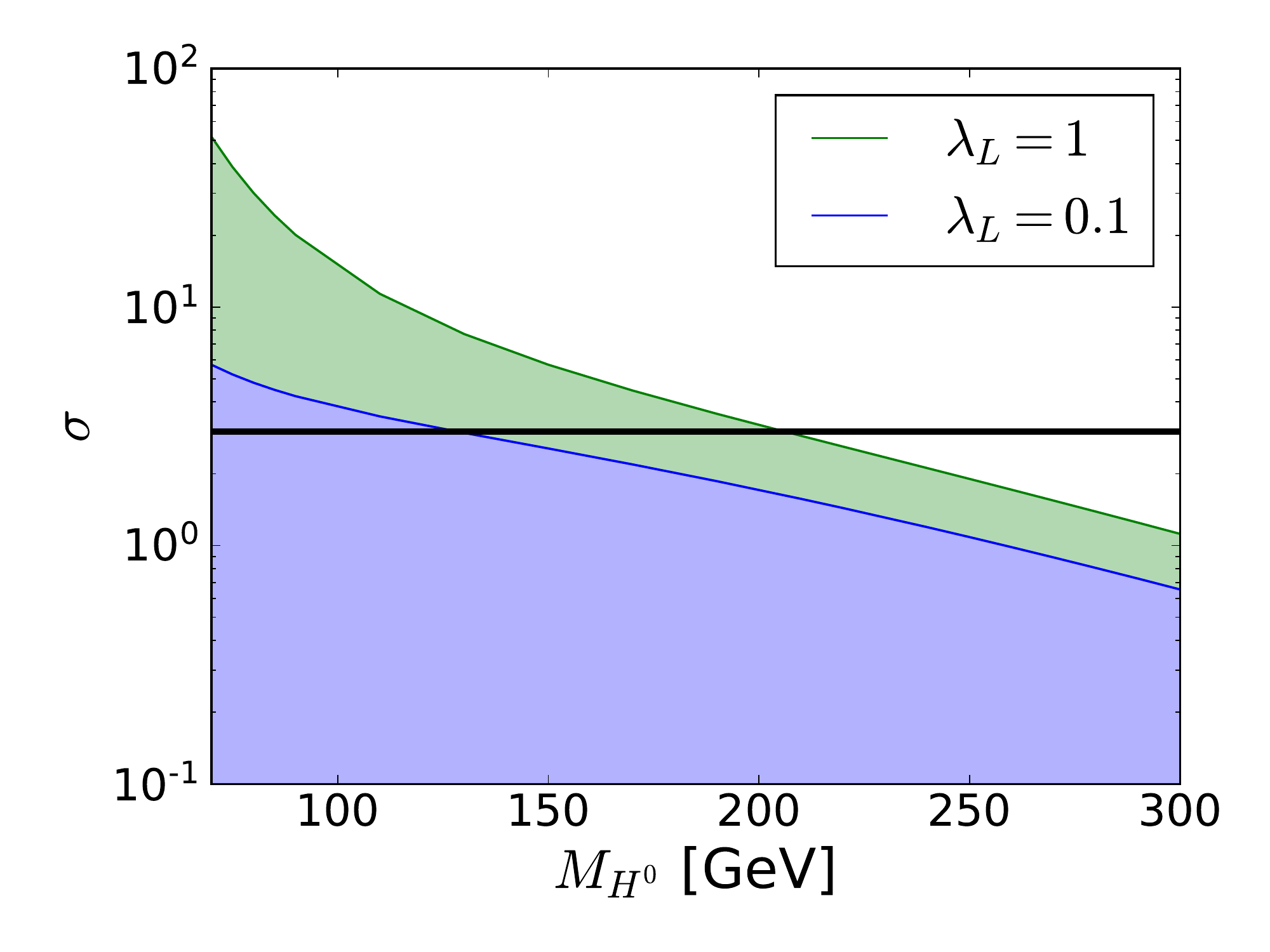} 
  \includegraphics[scale=0.43]{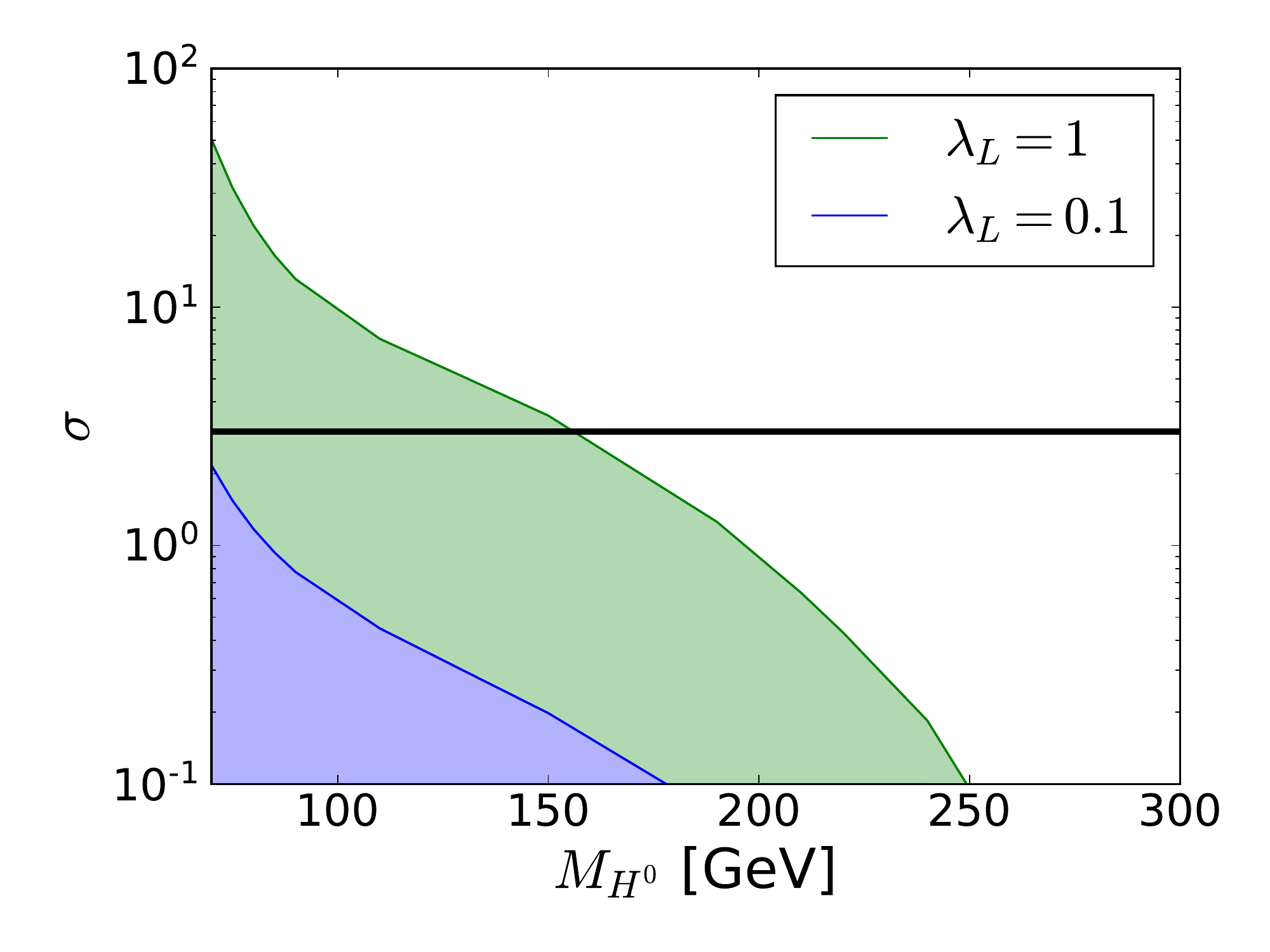}
  \caption{Significance for DM candidate discovery  for a luminosity of $3000\ \text{fb}^{-1}$. The charged scalar mass is $750\ \text{GeV}$ ($250\ \text{GeV}$) in the left  (right) panel }
  \label{fig:f2}
\end{figure}

Using the cuts developed in table~\ref{tab:cutflow}, we show the significances as a function of $M_{H_0}$ in figures~\ref{fig:f1} and~\ref{fig:f2} for 30 and 3000 fb$^{-1}$ luminosities and two different values of $\lambda$. In table~\ref{tab:cutflow} we have quoted only statistical uncertainties, however in our results we have considered a higher uncertainty of 30\% over signal and background yields to have a more realistic approach including other sources of uncertainties (e.g. PDF, cross section, ...). We find that larger values of $M_{H^{\pm}}$ and $\lambda_L$ produce larger significances for VBF analysis due to large production cross-sections.  For $\lambda_L\sim 1$, the 3$\sigma$ reach can go up to $M_{H^0}\sim$200 GeV for 3000 fb$^{-1}$ luminosity. 

\begin{figure}
  \centering
  \includegraphics[scale=0.5]{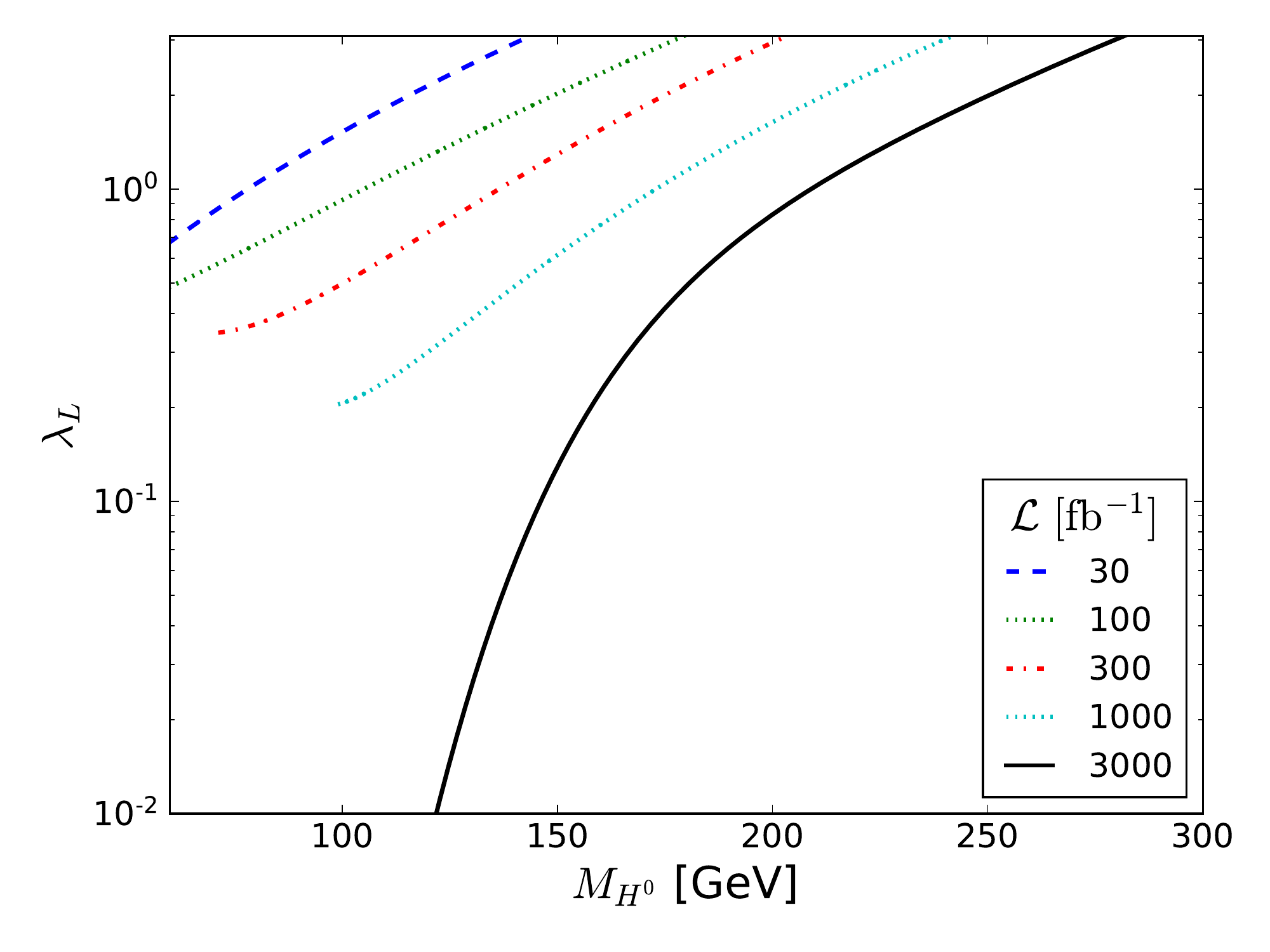}

  \caption{3 $\sigma$ reach for $\lambda_L$ vs  $M_{H^0}$ parameter space for various luminosities in $\text{fb}^{-1}$. }
  \label{fig:lc}
\end{figure}

In figure~\ref{fig:lc}, we show the 3$\sigma$ reach of our VBF cuts in the $\lambda_L$ and $M_{H^0}$ parameter space for various luminosities ranging from 30 to 3000 fb$^{-1}$. We maximized the production cross-sections by choosing $M_{H^{\pm}}$ appropriately. We can see that the reach for the   parameter space is substantial. The 3$\sigma$ reach of the  parameter space of  can go up to 280 GeV of $M_{H^0}$ for larger values of $\lambda_L$. For $\lambda_L\sim 10^{-2}$, the 3$\sigma$ reach for the DM mass goes up to  $M_{H^0}\sim$125 GeV. 

The IDM can be constrained using monojet-type searches at the LHC. However the reach of the VBF search is better than the monojet search.  In Table~\ref{tab:monjet-VBF}, we compare the significances of  monojet search using the cuts as given  in~\cite{Aaboud:2016tnv, Belyaev:2016lok}  with the VBF search strategy as developed in this paper at 20 fb$^{-1}$. In the table, $S_{\text{MJ, VBF}}$, $B_{\text{MJ, VBF}}$ and $\sigma_{\text{MJ,VBF}}$ are signal, background events and significances for monojet and VBF selection cuts respectively where we find that the VBF searches are more effective in probing the IDM models. 

\begin{table}
\centering
\begin{tabular}{|c|c|c|c|}\hline
$\lambda_L$& $S_{\text{MJ}}$ & $B_{\text{MJ}}$ & $\sigma_{\text{MJ}}$ \\\hline
0.01 & 3 &  & 0.015 \\ 
0.1 & 3.7 & 134044 & 0.02 \\ 
1.0 & 31 & & 0.15\\ \hline 
\end{tabular}\hspace{1cm}
\begin{tabular}{|c|c|c|c|}\hline
$\lambda_L$& $S_{\text{VBF}}$ & $B_{\text{VBF}}$ & $\sigma_{\text{VBF}}$ \\\hline
0.01 & 4.6 &  & 0.21 \\
0.1 & 7.5 & 476 & 0.35 \\
1.0 & 25 & & 1.1\\  \hline
\end{tabular}

\caption{Sensitivities for $M_{H^+}=M_{A^0}=750\ $GeV and $M_{H_0}=110\ $GeV for several  $\lambda_L$ values, with an integrated luminosity of $20$~fb$^{-1}$.}
\label{tab:monjet-VBF}
\end{table}

If $M_{H^+}$ is not so large ($<500$ GeV), $H^+$ can be produced efficiently with larger cross-section which then subsequently decay into $H^0$. In such parameter space, Drell-Yan type cuts can be useful.
In table~\ref{tab:DY}, we show  some Benchmark points where we vary  $H^+$, $H^0$ masses and $\lambda_L$ to see  effects on $\sigma(pp \to H^{0}H^{0} j j )$ where we find that if the charged Higgs mass is 
smaller, then the cross-section gets larger.  
 
\begin{table}[h]
\centering
\begin{tabular}{|c|c|c|c|c|c|}\hline
Benchmark & $m_{H^{+}}$ [GeV] & $m_{A^{0}}$ [GeV] & $m_{H^{0}}$ [GeV] & $\lambda_L$ & $\sigma(pp \to H^{0}H^{0} j j )$[pb]  \ \\\hline
BP1   & 200 & 189.5 &   65 & 0.009 & 0.067 \\ \hline
BP2  &  500 &  494 &   65 &  0.009 &  0.009\\ \hline
BP3 &  750 &  750 &   65 &  0.009 &  0.008\\ \hline
BP4 &  750 &  750 &   110 &  0.009 &  0.005\\ \hline
BP5 &  750 &  750 &   65 &  0.500 &  0.274\\ \hline
\end{tabular}
\caption{Benchmark scenarios for the comparison}
\label{tab:DY}
\end{table}

We then apply VBF and D-Y type cuts on all the scenarios where the D-Y type cuts are defined in~\cite{Poulose:2016lvz}: $N(j) = 2$, $N(b) = 0$, $N(l) = 0$, 
$p_T^{miss} > 260$ GeV, $p_T (j_1) > 120$ GeV , $p_T (j_2) > 90$ GeV, 
$75\,{\rm GeV} \leq M_{j_1j_2}\leq90$ GeV, $\Delta R_{j_1j_2} < 1.8$.
We show our results in table~\ref{tab:DY-VBF} where we compare these cuts. In the table, $S_{\text{DY}, \text{VBF}}$, $B_{\text{DY}, \text{VBF}}$ and $\sigma_{\text{DY},\text{VBF}}$ are signal, background events and significances for D-Y and VBF selection cuts respectively. We find that for $M_{H^+}\geq 500$ GeV, the significance becomes much better for VBF analysis.
 
\begin{table}
\centering
\begin{tabular}{|c|c|c|c|c|}\hline
Benchmark & $B_{\text{DY}}$ & $S_{\text{DY}}$ &$\sigma_{\text{DY}}$ \ \\\hline
BP1 &  & 148&1.59 \\
BP2 &  & 68  & 0.73\\
BP3   & 8500 & 15 &  0.16\\
BP4   &  & 10  & 0.11\\ 
BP5   &  & 85   & 0.92\\ \hline
\end{tabular}\hspace{1cm}
\begin{tabular}{|c|c|c|c|c|}\hline
Benchmark & $B_{\text{VBF}}$ &  $S_{\text{VBF}}$  &  $\sigma_{\text{VBF}}$\ \\\hline
BP1   &  &  38&0.25 \\ 
BP2 &  &  198 &1.28\\
BP3   & 23809 &  309& 1.99\\
BP4   &  &  234 &1.51\\
BP5   &  &  3579 &21.6\\ \hline
\end{tabular}
\caption{Comparison between our analysis and the one implemented in Ref.~\cite{Poulose:2016lvz},
with an integrated luminosity of $1000$~fb$^{-1}$.  }
\label{tab:DY-VBF}
\end{table}

\section{Conclusions}
\label{sec:conclusions}

In this paper we utilize VBF search strategy to probe the parameter space of the inert doublet model where $H_0$ is a dark matter candidate. We probe the parameter space of the model at the LHC without applying constraints from the annihilation rate and the data from the direct and indirect experiments,  since these constraints can be relaxed in various cosmological scenarios. 

The VBF search is centered around the requirement of two high $p_T$ jets  which located in different hemispheres of the detector,  $\eta(j_{1})\times\eta(j_{2})<0$, which are also well separated in the pseudorapidity and their invariant mass is larger than for a couple of non-VBF jets. We further optimize the VBF cuts and found that $p^{miss}_T>$180 GeV, $p_T(j_{1(2)})>100 (50)$ GeV, $|\Delta\eta|>4.2$, $M(j_1,j_2)>1$~TeV, $N(j)\geq$ 2, $\eta(j_1)\times\eta(j_2)<$0 provide the largest significance. The dominant background arises from $Z$+jets and we showed   the signal and background distributions after each of these cuts. We  showed that VBF search has a better reach  when compared with the monojet searches. We also showed that for  $M_{H^{+}}<500$ GeV, D-Y type of cuts provide better significances while VBF cut  works much better for heavier  $M_{H^{+}}$.  We showed the 3$\sigma$ reach of the  $\lambda_L-M_{H^0}$ parameter space of of the model using VBF cuts for various luminosities and found that the reach can go up to 280 GeV of Higgs mass for a luminosity of 3000 fb$^{-1}$. 

\section{Acknowledgments}

B. D. acknowledges support from  DOE Grant de-sc0010813.
D. R. acknowledges support from  the Grants Sostenibilidad-GFIF and CODI-IN650CE, and  COLCIENCIAS through the Grant No. 111-565-84269.
J. D. Ruiz-Álvarez gratefully acknowledges the support of COLCIENCIAS, the Administrative Department of Science, Technology and Innovation of Colombia.

\bibliographystyle{apsrev4-1long} 
\bibliography{susy}

\end{document}